%% file: main.tex
\documentclass[9pt,twocolumn]{article} 
\usepackage{graphicx}
\usepackage{textcomp}
\usepackage[a4paper, total={6in,10in}]{geometry}
\usepackage{appendix}

\title{Towards high-energy few-cycle optical vortices with minimized topological charge dispersion}
\date{}

\begin{document}

\twocolumn[
\maketitle

\noindent Federico J. Furch$^{1,*}$, Gunnar Arisholm$^2$
\\
\textit{$^1$Max Born Institute for Nonlinear Optics and Short Pulse
  Spectroscopy, Max-Born-Strasse 2a, 12489 Berlin, Germany}
\\
\textit{$^2$Norwegian Defence Research Establishment (FFI), PO Box 25, 2027 Kjeller, Norway}
\\
\textit{$^*$Corresponding author: furch@mbi-berlin.de}
\\

\textit{ A simple approach to generate high-energy few-cycle optical vortices with minimized topological charge dispersion is introduced. By means of numerical simulations it is shown that, by leveraging the intrinsic properties of optical parametric chirped pulse amplification (OPCPA), clean transfer of topological charge from a high energy narrowband pump pulse to a broadband idler is feasible under certain particular conditions, enabling the generation of high-energy few-cycle vortex pulses with extremely low topological charge dispersion. }
\vspace*{4ex}
]

In 1989 Coullet et al., proposed the existence of optical vortices in laser cavity modes \cite{COULLET1989}. A few years later Allen and co-workers demonstrated that Laguerre-Gaussian (LG) spatial amplitude distributions with azimuthal index different from zero, which constitute one type of optical vortex, carry orbital angular momentum (OAM) \cite{Allen1992}. Since then, optical vortices have emerged as powerful tools across a wide range of applications including but not limited to, optical trapping and manipulation, optical communications, high-dimensional quantum cryptography, laser micromachining and microscopy \cite{Shen2019}. Moreover ultrashort optical vortices with high energy have been proposed for diverse applications such as hollow plasma drilling \cite{Wang2020} or acceleration of attosecond electron slices \cite{jiang2021}. In particular, it has been demonstrated that high harmonic generation driven by ultrashort IR fields carrying a topological charge can produce extreme ultraviolet light with OAM \cite{Zuerch2012,Gariepy2014,Gauthier2017}. These novel laser fields can lead, among other interesting effects, to spatially-dependent modified selection rules during single photon ionization \cite{Afanasev2018,Picon2010} and enable OAM-induced x-ray dichroism \cite{vanVeenendaal2007}.
 
 An optical vortex is characterized by a point of zero intensity and an azimuthal phase variation that integrates to $2\pi m$ on a closed path around that point, for a nonzero integer $m$. While vortices can appear randomly in speckle patterns, the term vortex beam describes a beam with a well-defined complex amplitude pattern and a single vortex in the center. This form of phase variation means that the wavefront is helical. $m$ is called the topological charge, and in the context of LG modes it is indicated by the azimuthal index of the mode. Although a variety of methods for generating vortex beams have been explored, most notably, a combination of cylindrical lenses \cite{Allen1992},  spiral phase plates \cite{Beijersbergen1994}, liquid-crystal-based spatial light modulators \cite{Brunet2014}, and holographic plates \cite{Heckenberg1992}, these approaches rely on imparting a specific spatial phase at a particular target wavelength. Since the spatial spiral phase varies for different wavelengths, implementation of these methods with broadband sources leads to what is known as topological charge dispersion, i.e., different content of topological charges for different wavelengths of the spectrum. Topological charge dispersion leads among other things to a non-preserving spatial distribution during propagation and complex spatio-temporal structures \cite{Piccardo2023}. 
 
A number of methods have been proposed and implemented in order to address the topological charge dispersion challenge. Bock et al. employed a diffractive-refractive element to obtain few-cycle vortices at a specific plane from Ti:Sapphire oscillator pulses \cite{Bock2013, Musigmann2014}. Tokizane et al. combined achromatic wave plates and a segmented half wave plate for broadband vortex beam generation \cite{Tokizane2009}. The same group further extended this approach, amplifying a broadband vortex beam produced in such way in a 2-stage optical parametric amplifier \cite{Yamane2012}. Amidst these methods, Atencia et al. developed a compound holographic optical element to generate achromatic vortices \cite{Atencia2013}, while Swartzlander designed achromatic spiral phase plates from two different glasses, providing a perfect topological charge at two wavelengths, akin to achromatic lenses \cite{Swartzlander2006}.

 With the notable exception of Naik et al., who introduced an elegant approach based on a Sagnac interferometer to generate achromatic vortices \cite{Naik2017}, these techniques often prove complex or challenging to implement, and/or incompatible with the generation of high-energy ultra broadband vortex pulses. An alternative to generating a wide-band optical vortex directly (i.e., transforming a field without OAM), is to use an optical parametric chirped pulse amplifier (OPCPA)\cite{DUBIETIS1992} to transfer the OAM of a narrow-band pump beam to a wide-band idler beam. The phase-sensitive nature of the parametric amplification process implies that the idler produced during the amplification of a weak signal wave carries the phase difference between the pump and signal waves. This also holds for helical wavefronts, so the difference in topological charge between the pump and the signal will be transferred to the idler. The concept of transferring the topological charge from the pump to the idler was originally demonstrated in an optical parametric amplifier (OPA) with continuous wave sources \cite{Caetano2002} and later implemented in pulsed narrowband optical parametric oscillators where the selective transfer of the topological charge of the pump to the signal or idler was demonstrated \cite{Yusufu2012}. Camper et al. demonstrated the generation of a high energy femtosecond optical vortex in the idler of a noncollinear OPA pumped by a vortex beam \cite{Camper2017}. This OPA featured relatively long pulses and consequently their analysis did not address topological charge dispersion. In this work it is shown that this concept can be extended to ultra broadband sources supporting few-cycle pulses. The topological charge content of the wide-band idler generated in an OPCPA pumped by a narrowband vortex is analyzed and it is shown that, under certain conditions, the topological charge dispersion can be minimized over the entire bandwidth.

\begin{figure}[!thb]
\centering
\fbox{\includegraphics[width=0.95\linewidth]{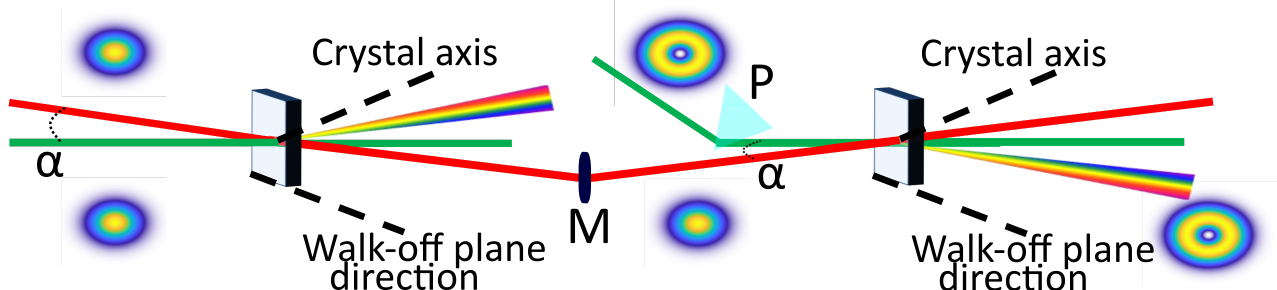}}
\caption{Proposed experimental scheme. The beam profiles illustrate Gaussian or Laguerre-Gaussian (LG) modes. The prism (P) illustrates an optical element introduced to achieve pulse-front matching. The LG mode of the output idler is also indicated. M represents a set of mirrors to transport/re-size the amplified signal into the second amplification stage.}
\label{fig:schemes}
\end{figure}

The system under study is the two-stage noncollinear OPCPA schematically shown in Fig.  \ref{fig:schemes}. The first stage had Gaussian beams for both pump and seed, and the second stage was pumped by an LG mode. The 3D numerical simulations of the parametric processes were performed utilizing the code Sisyfos (SImulation SYstem For Optical Science) \cite{Arisholm1997,Arisholm1999,Sisyfos}. The code allows simulating second-order frequency mixing processes including the effects of pump depletion, back-conversion, walk-off, dispersion, diffraction, parasitic effects, and thermal effects. The coupled-wave equations in three dimensions are solved in Fourier space. This code has been utilized before to simulate OPCPAs under realistic conditions \cite{Schlup2004,Arisholm2004,Giree2017}.

All the simulations presented in this work are based on noncollinear type-I phase-matching in BBO crystals, specifically with a noncollinear angle $\alpha = 2.5 ^{\circ}$. The first stage was under the walk-off compensation (WOC) geometry ($\theta_{signal} = 0.4702$ rad, $\theta_{pump} = 0.4265$ rad, with $\theta$ the angle between the crystal axis and the k-vector of the signal or the pump). The WOC geometry minimizes space-time couplings in the amplified signal but suffers from parasitic second harmonic of signal and idler \cite{Bromage2011} (taken into account in the simulations). Meanwhile, the second stage was under tangential phase-matching or non-walk-off compensation geometry ($\theta_{signal} = 0.3829$ rad, $\theta_{pump} = 0.4265$ rad), which is better suited for large beams for which the lateral walk-off between the beams is small compared to the beam sizes. The pulse parameters  
employed herein are similar to a near-infrared OPCPA system developed at the Max Born Institute \cite{Furch2017, Witting2018, Furch2022}. A measured pulse spectrum from a Ti:Sapphire oscillator was utilized to build the input seed pulses, stretched by applying positive group delay dispersion (800 fs$^2$), and energy of 1 nJ. The pump pulses were assumed to have a Gaussian temporal shape with Fourier-Transform limited duration of 1 ps, centered at 515 nm.

It was assumed that the pump beam is transformed into an LG mode by means of a spiral phase plate, which are nowadays commercially available for a target wavelength of 515 nm. Spiral phase plates provide high damage threshold and are easy to implement experimentally. For the simulations shown in this work it was assumed that the spatial distribution of the pump field is described by an LG mode with a topological charge of $m=1$ and no topological charge dispersion. It is important to remark that it was also assumed that in the second stage pulse front matching was implemented \cite{Kobayashi2002} (illustrated by the prism (P) in Fig. \ref{fig:schemes}). The results can degrade substantially without pulse front matching (see \textbf{Supplementary material}).

\begin{figure}[!htb]
\centering
\fbox{\includegraphics[width=0.95\linewidth]{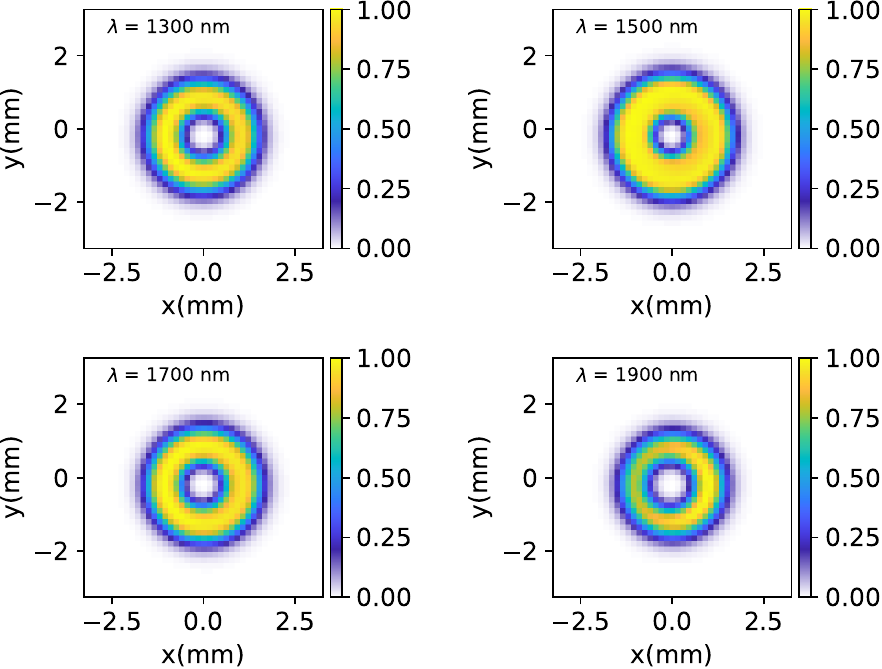}}
\caption{Frequency resolved idler beam profiles at selected wavelengths of the spectrum. The color scale indicates the normalized intensity}
\label{fig:idler_beam_case1}
\end{figure}
\begin{figure}[!htb]
\centering
\fbox{\includegraphics[width=0.95\linewidth]{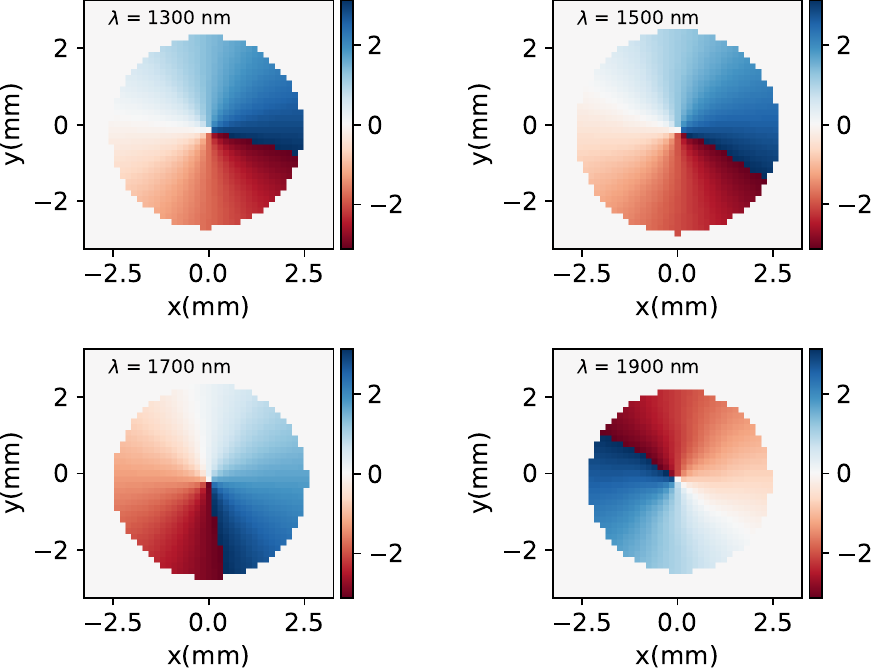}}
\caption{Frequency resolved idler spatial phase profiles at selected wavelengths of the spectrum. The color scale indicates phase in radians}
\label{fig:idler_phase_case1}
\end{figure}

In the first OPCPA stage the pump pulse energy was 0.4 mJ and the peak intensity incident on the crystal was 97 GW/cm$^2$. The pump and signal beams in the first OPCPA stage had matching sizes (700 \textmu{}m diameter $1/e^{2}$). The signal pulses were amplified in a 1.5 mm thick crystal to 14.9 \textmu{}J. These amplified signal pulses were utilized to seed the second amplification stage, pumped by an LG mode with a topological charge of $m=1$, pulse energy of 5 mJ and peak intensity of 50 GW/cm$^2$ incident on a 1.2 mm thick BBO crystal. In this case the signal beam diameter was 4.8 mm, approximately 20$\%$ larger than the pump. At the output of this second OPCPA stage the idler carries the same topological charge as the pump. This is illustrated in Fig. \ref{fig:idler_beam_case1} and Fig. \ref{fig:idler_phase_case1} showing frequency resolved beam profiles and spatial phases at selected wavelengths. The beam profiles show a clear annular shape with the amplitude vanishing in the center. The noncollinear geometry and spatial walk-off cause an uneven spatial intensity distribution (left/right asymmetry), especially towards the edges of the spectrum. The phase profiles show in all cases illustrated in Fig. \ref{fig:idler_phase_case1} the characteristic azimuthal phase variation. The topological charge dispersion of the idler was quantified applying the method of Wang et al., who studied the effects of parametric amplification over the topological charge content of a broadband optical vortex used as seed of the amplifier \cite{Wang2023}. In order to calculate the topological charge content they expanded the spatial part of the laser electric field in helical harmonics. Here it is assumed that this is done at a particular frequency component:

\begin{equation}
E(r,\phi) = \frac{1}{\sqrt{2\pi}}\sum_{m=-\infty}^{\infty}a_m(r)e^{im\phi}
\label{eq:expansion_E}
\end{equation}

where the functions $a_m(r)$ can be calculated as

\begin{equation}
a_m(r) = \frac{1}{\sqrt{2\pi}}\int_{0}^{2\pi}E(r,\phi)e^{-im\phi}d\phi
\label{eq:am}
\end{equation}

To obtain a coefficient corresponding to a particular topological charge $m$ the absolute value square of the function $a_m(r)$ is integrated:

\begin{equation}
b_m = \int_{0}^{\infty}|a_m(r)|^2 rdr
\label{eq:bm}
\end{equation}

Finally, the proportion of a particular coefficient relative to all others in the expansion can be simply calculated as

\begin{equation}
P_m = \frac{b_m}{\sum_{n=-\infty}^{\infty}b_n(r)}
\label{eq:Pm}
\end{equation}

\begin{figure}[!bht]
\centering
\fbox{\includegraphics[width=0.9\linewidth]{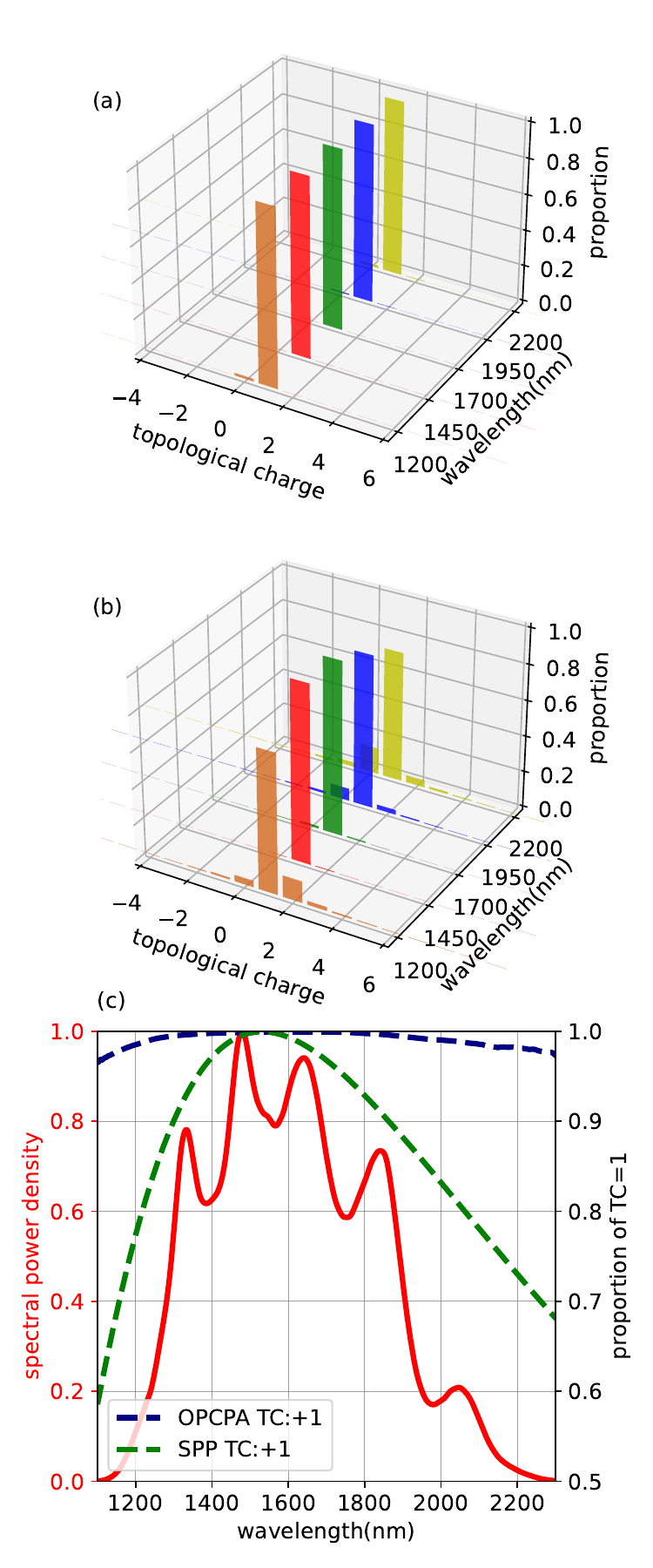}}
\caption{Topological charge dispersion in the idler. (a) Relative topological charge content ($P_m$ coefficients) at selected wavelengths across the idler spectrum. (b) Relative topological charge content assuming a spiral phase plate (SPP) has been utilized to convert the broadband Gaussian field into a Laguerre-Gaussian (c) Idler spectrum (red), relative proportion of topological charge $m=1$ across the spectrum for the optical vortex generated during parametric amplification (dashed blue) or utilizing an SPP (dashed green line).}
\label{fig:idler_TC_case1}
\end{figure}

Equation \ref{eq:Pm} offers insight into the topological charges forming the spatial profile at a particular frequency component. Figure \ref{fig:idler_TC_case1} (a) shows the distribution of topological charges calculated at selected wavelengths over the idler spectrum. It is observed that the topological charge of the pump has been transferred to the idler with remarkable purity. In addition, Fig. \ref{fig:idler_TC_case1} (c) shows the idler spectrum (red) and illustrates the proportion of topological charge $m=1$ across the entire idler spectrum (dashed blue). For comparison Fig. \ref{fig:idler_TC_case1} (b) shows the distribution of topological charge in the case in which a fused silica spiral phase plate designed for 1600 nm is utilized to produce a vortex beam with topological charge $m=1$ for an initial Gaussian spatial distribution. In addition, the dashed green line in Fig. \ref{fig:idler_TC_case1} (c) illustrates the proportion of topological charge $m=1$ across the spectrum in this case. Converting a broadband OPCPA seed into a vortex beam using a spiral phase plate (before amplification) is an approach that has been reported in the literature \cite{Wangjun2020,Wang2023}, and in particular the effect of saturation over the topological charge dispersion has been observed.   

Under the amplification conditions previously described the signal field was further amplified to an energy of 1.02 mJ and the resulting idler had a pulse energy of 0.59 mJ. This corresponds to a total extraction efficiency (considering both, signal and idler) of 32.2 $\%$ and represents a reasonable compromise between efficient energy extraction and minimization of space-time couplings that degrade the peak intensity of the amplified pulses \cite{Giree2017}. The spectrum of Fig. \ref{fig:idler_TC_case1} (c) supports a Fourier-transform limited pulse duration of 8.5 fs, corresponding to approximately 1.67 optical cycles at the carrier frequency 195.9 THz (1531 nm) (see \textbf{Supplementary material}). The angular dispersion of the idler can be compensated with established methods \cite{Kobayashi2002, Heiner2018} producing high energy, sub-2 cycle optical pulses in the short-wave infrared with a clean topological charge content. Alternatively, a controlled angular dispersion could be introduced in the signal beam seeding the second stage to obtain an angular-dispersion-free idler \cite{Huang2012}.

The scenario proposed here represents just one example of many different variations that can be implemented around the same idea. Other approaches include seeding a last OPCPA stage with the idler (with the appropriate angular dispersion) from a previous stage and pumped by an LG mode to obtain an angular-dispersion-free signal with controlled topological charge, combining optical vortices in both pump and seed fields (signal or idler) to control the topological charge of the third field generated during amplification, or designing a multi-stage, high energy OPCPA system in which the last amplification stage is pumped by an optical vortex to scale up the pulse energy as already proposed in \cite{Camper2017} for a femtosecond OPA. Additionally, similar scenarios based on sum frequency generation \cite{Pires2019} can be implemented. Furthermore, nonlinear propagation in thin glass plates could be utilized to push the pulse duration towards the single-cycle limit \cite{Lu2018,Chen2022}. Generating few- and near-single-cycle signal and idler pulses with controlled topological charge can be exploited in ultrafast two-color pump-probe spectroscopic techniques and XUV attosecond pulse generation with controlled OAM \cite{Gariepy2014,Gauthier2017}. 

In conclusion, this study demonstrates the remarkable capability to transfer the topological charge of a narrowband, high-energy laser pulse with exceptional fidelity to broadband idler pulses during optical parametric chirped pulse amplification. The proposed approach opens the door for the generation of high-energy few-cycle optical vortices with fine control over the topological charge. Follow up work will focus on the experimental demonstration of the proposed idea.

\bigskip
\noindent \textbf{Funding.} European Union Horizon 2020 program Laserlab Europe
(871124).
\\

\noindent The authors thank Achut Giree for useful discussions.

\bibliographystyle{ieeetr}
\bibliography{biblio_main}

\newpage

\renewcommand{\thesection}{\Alph{section}}
\setcounter{figure}{0}
\renewcommand\thefigure{S.\arabic{figure}}

\onecolumn
\appendix

\input{supplementary}

\end{document}

%% file: supplementary.tex
\section*{Supplementary material}
\vspace*{2ex}
\textit{In the main manuscript it has been shown that the topological charge of an energetic narrowband optical vortex pumping an optical parametric chirped pulse amplification (OPCPA) stage can be transferred to the generated idler with high fidelity over the whole resulting ultra broad spectrum which supports few-cycle pulse duration. But amplification of extremely broadband pulses requires a noncollinear phase-matching geometry, which inherently introduces space-time couplings \cite{Bromage2010, Bromage2011, Zaukevicius2011, Giree2017}. In this additional material the influence of pulse front matching on the fidelity of the topological charge transferred is briefly explored, complementing the results presented in the main manuscript. In addition, the compressibility of the idler pulses is corroborated by compensating the group delay dispersion.}
\vspace*{2ex}

\section{Influence of pulse front matching}

One of the well studied and well understood effects of the noncollinear phase-matching geometry in OPCPAs is the emergence of spatial chirp in the amplified signal. This effect originates in the effective tilted gain experienced by the seed (signal) field due to the angle between the pulse fronts of the signal and pump (see for example Fig. 1 in \cite{Zaukevicius2011}). As a result, the idler not only experiences angular dispersion, essential for satisfying phase-matching over a broad spectral range, but also suffers from additional spatial chirp. One way of visualizing the spatial chirp is by observing the frequency resolved beam profile at particular frequencies over the spectrum. The center of gravity of the beam moves with frequency in the walk-off plane of the OPCPA stage. Of course, in the case of the idler, this also happens due to angular dispersion. 

In order to explore this effect more closely in the case of OPCPAs pumped with an optical vortex a set of additional simulations were performed. In all of them, a noncollinear stage was simulated in the non-walk-off compensation (NWOC) geometry and a seed almost identical to the seed in the first stage described in the main manuscript (i.e., same spectrum, chirp, energy). The pump field was characterized by a Laguerre-Gaussian (LG) mode with a topological charge $m=1$, with the same spectrum and pulse duration of the pump fields utilized in the simulations of the main manuscript, and pulse energy of 5 mJ. The pump peak intensity incident on the 2 mm thick BBO crystal was 50 GW/cm$^2$ and the seed beam size was slightly larger ($\approx$ 20$\%$) than the pump beam. In the choice of input parameters it was intended to minimize other effects inducing space-time couplings such as saturation (combined extraction efficiency of signal plus idler below 1$\%$) and the lateral translation of the beams during propagation. 

\begin{figure}[!hbt]
\centering
\fbox{\includegraphics[width=.57\linewidth]{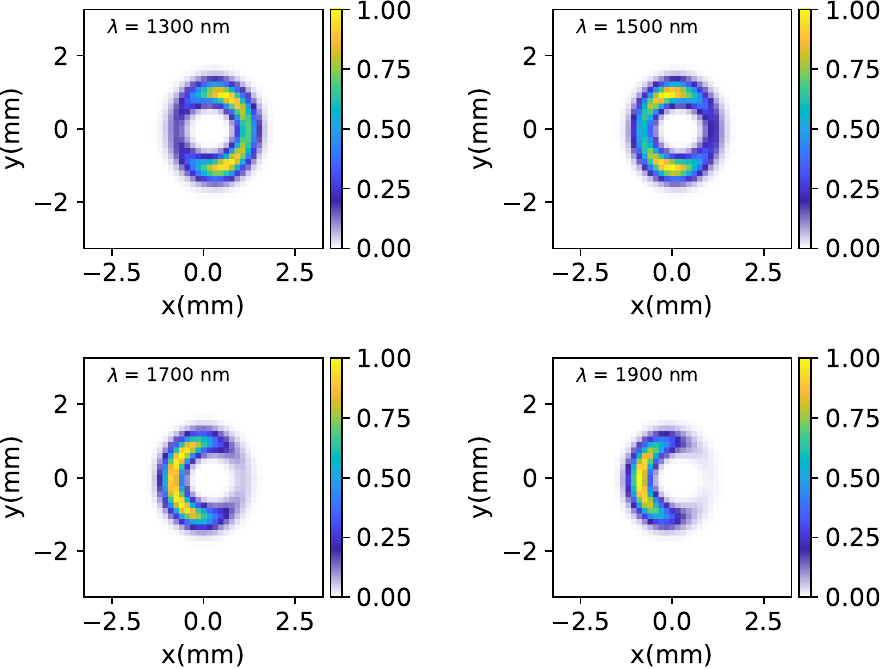}}
\caption{Frequency resolved beam profiles at selected frequencies of the idler spectrum. The pulse fronts are not tilted, that is, the pulse front of each beam is perpendicular to the corresponding k-vector.}
\label{fig:beams_no_pfm}
\end{figure}
Figure \ref{fig:beams_no_pfm} shows the frequency resolved beam profiles for the resulting idler. In the case of a vortex beam pumping the amplifier, the spatial chirp leads to an interesting effect, namely the beam profile is a ring only around the center of the spectrum, while towards the edges of the spectrum it turns into a half-ring shape, with opposite halves of the ring-like mode appearing in opposite ends of the spectrum.

\begin{figure}[!htb]
\centering
\fbox{\includegraphics[width=.6\linewidth]{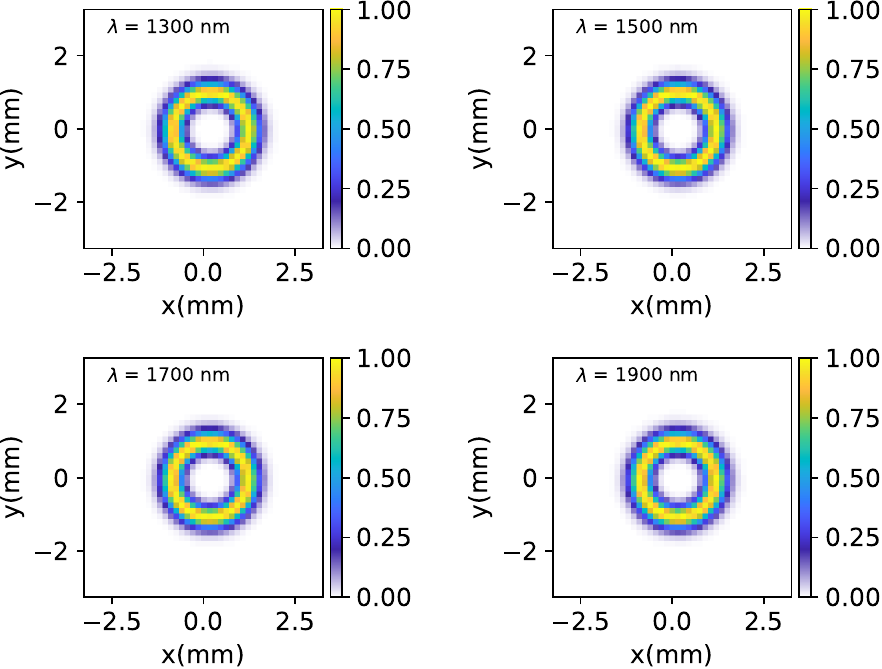}}
\caption{Frequency resolved beam profiles at selected frequencies of the idler spectrum. The pulse front of the pump is tilted to be parallel with the pulse front of the signal (pulse-front matching).}
\label{fig:beams_pfm}
\end{figure}

In stark contrast, the beam profiles at different frequencies is relatively homogeneous when pulse front matching \cite{Kobayashi2002} is implemented. Fig. \ref{fig:beams_pfm} shows the results of the simulation for the case of pulse-front matching between signal and pump. 

\begin{figure}[!htb]
\centering
\fbox{\includegraphics[width=.6\linewidth]{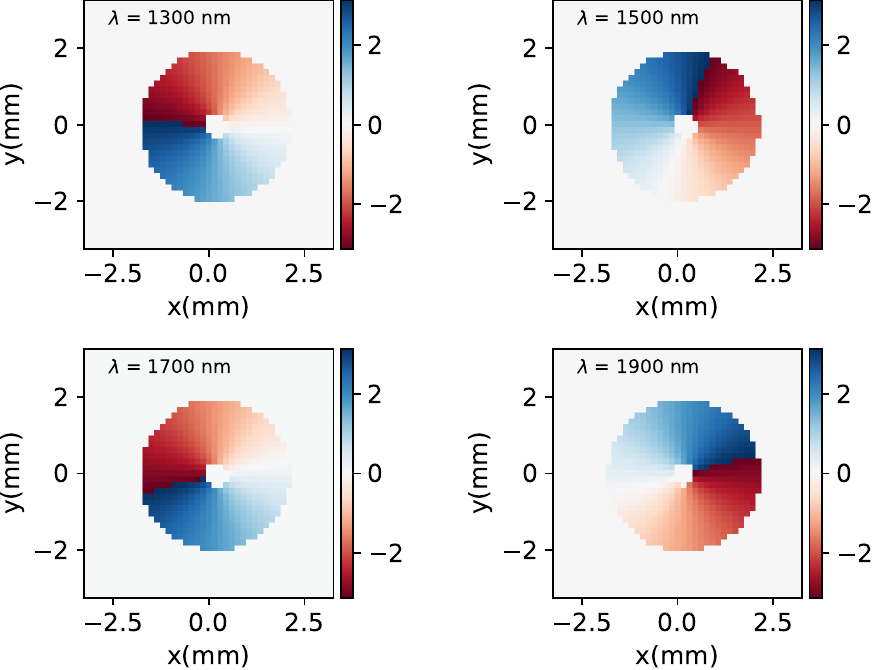}}
\caption{Frequency resolved phase profiles at selected frequencies of the idler spectrum. Pulse-front matching between signal and pump has been implemented.}
\label{fig:phases_pfm}
\end{figure}
 
\begin{figure}[!hbt]
\centering
\fbox{\includegraphics[width=.6\linewidth]{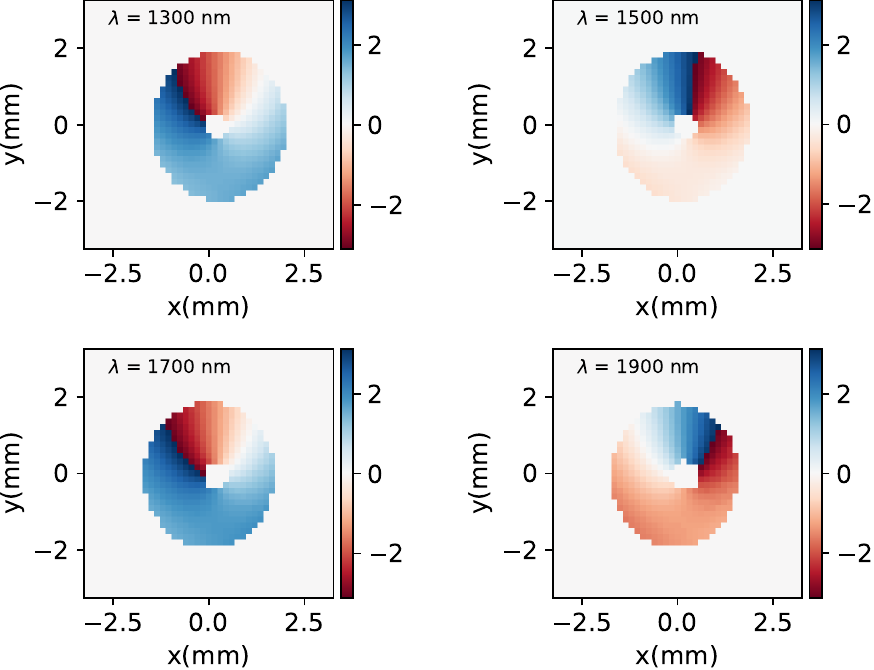}}
\caption{Frequency resolved phase profiles at selected frequencies of the idler spectrum. The pulse fronts are not tilted. The pulse front of each beam is perpendicular to the corresponding k-vector.}
\label{fig:phases_no_pfm}
\end{figure}

The frequency resolved phase profiles are also affected. In this case the characteristic homogeneous azimuthal phase variation that can be observed when the pulse fronts are matched (Fig. \ref{fig:phases_pfm}) acquires a top/bottom asymmetry when the pulse-fronts of signal and pump subtend an angle (Fig. \ref{fig:phases_no_pfm}). The azimuthal variation of the phase is preserved, but the change is not constant (or linear) over the azimuthal coordinate. As it can be observed in Fig. \ref{fig:phases_no_pfm}, the phase variation is faster in the upper half-plane ($y>0$). Interestingly this top/bottom asymmetry reverses when the sign of the topological charge changes or when the phase-matching geometry changes (walk-off compensation or non-walk-off compensation). A detailed investigation of these effects is outside the scope of the present work and will be addressed in future work.

\begin{figure}[!htb]
\centering
\fbox{\includegraphics[width=.6\linewidth]{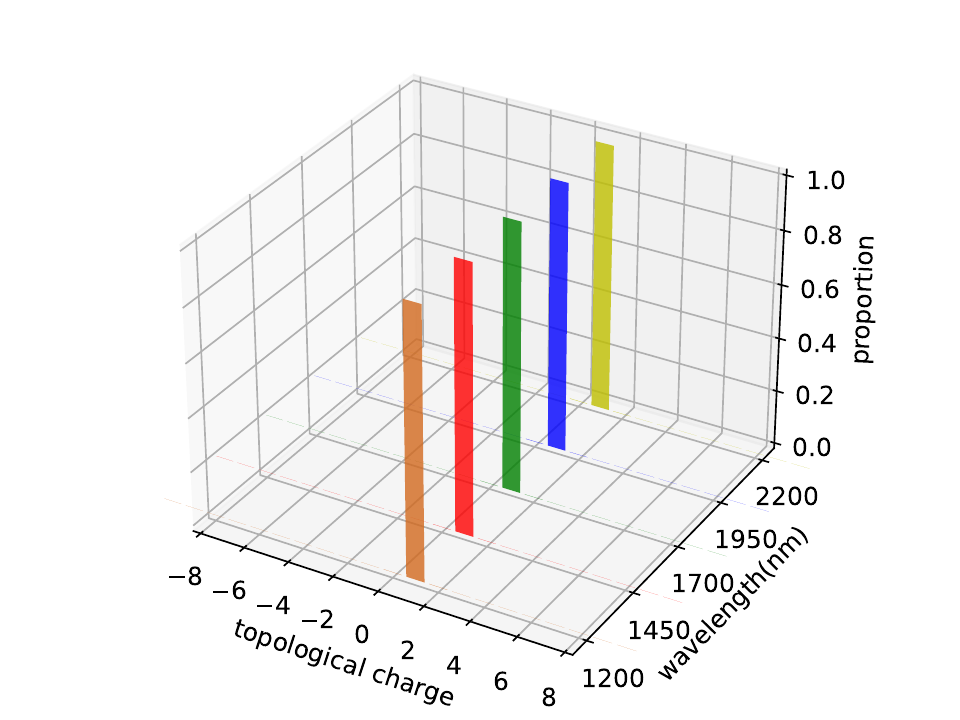}}
\caption{Relative topological charge content at selected wavelengths of the idler spectrum for an OPCPA stage in which pulse-front matching between signal and pump has been implemented.}
\label{fig:TC_pfm}
\end{figure}
 
\begin{figure}[!htb]
\centering
\fbox{\includegraphics[width=.6\linewidth]{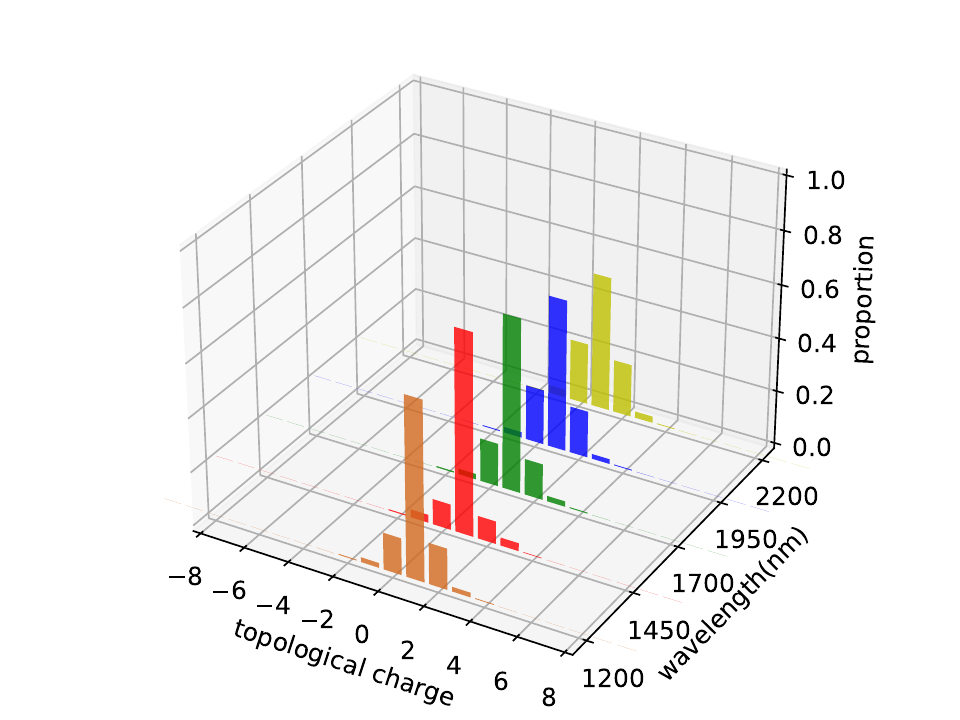}}
\caption{Relative topological charge content at selected wavelengths of the idler spectrum for an OPCPA stage in which the pulse-fronts of signal and pump subtend an angle equal to the noncollinear angle.}
\label{fig:TC_no_pfm}
\end{figure}

The consequence of the effects shown in Fig. \ref{fig:phases_no_pfm} and Fig. \ref{fig:beams_no_pfm} is a loss of fidelity in the transfer of topological charge during the amplification process. By simple inspection of equations 1-4 in the main manuscript, it is clear that the combination of the top/bottom asymmetry in the phase profiles and the left/right asymmetry in the amplitude (beam profiles) result in a broader distribution of topological charge. In other words, additional modes are necessary to describe the resulting spatial distribution. Figures \ref{fig:TC_pfm} and \ref{fig:TC_no_pfm} show the relative topological charge content for the case of pulse-front matching and when no pulse-front matching is implemented respectively.

\newpage
\section{Pulse compression to few-cycles}
\begin{figure}[!ht]
\centering
\fbox{\includegraphics[width=0.9\linewidth]{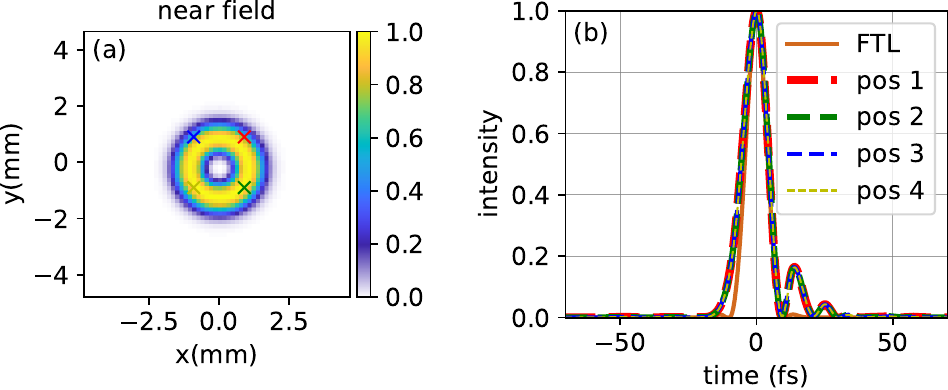}}
\caption{(a) Near field beam profile of the idler. The crosses correspond to the positions from which the spectral phase was extracted. (b) Spatially integrated pulse shapes corresponding to second-order phase compensation with the phases extracted at the positions marked in (a).}
\label{fig:pulses}
\end{figure}

In order to confirm compressibility to few-cycles, the idler pulses corresponding to the simulation shown in the main manuscript have been numerically compressed by imposing a global quadratic spectral phase across the whole spatial distribution. First, the spectral phase was extracted at a particular point of the near field distribution and approximated by a second order polynomial. Subsequently this second order phase was subtracted over the whole spatio-spectral distribution and the spatio-temporal distribution calculated by applying the inverse Fourier transform. Finally, the pulse shape was integrated over the entire spatial distribution ($\int |E(x,y,t)|^2 dxdy$). The procedure was repeated at four different positions on the near field distribution (marked with crosses in Fig. \ref{fig:pulses} (a)). The resulting pulse shapes corresponding to the four positions marked in \ref{fig:pulses} (a) are shown in \ref{fig:pulses} (b), together with the Fourier-transform limited (FTL) pulse.

The remarkable agreement between the pulse shapes obtained by extracting the phase from four different positions across the spatial distribution shows that the spatio-temporal distortions are low, as expected from the moderate extraction efficiencies and the pulse-front matching scheme. In all cases the pulse duration obtained by compensating the second-order of the spectral phase amounts to 10.2 fs, corresponding to 2 optical cycles, while the Fourier-limited pulse duration of 8.5 fs corresponds to 1.67 cycles. Compensation of the spectral phase up to third order leads to a pulse duration only $0.6 \%$ longer than the Fourier-transform limit (not shown in Fig. \ref{fig:pulses}).
